\documentclass[preprint,prd,nofootinbib,tightenlines]{revtex4}
\usepackage{graphicx}
\usepackage{bm}

\begin{document}
\baselineskip=15pt \parskip=5pt

\vspace*{3em}

\preprint{}

\title{\boldmath Rare $B$ Decays with a HyperCP Particle of Spin One}
\author{Sechul Oh}
\email{scoh@phys.ntu.edu.tw}
\affiliation{Department of Physics and Center for Theoretical Sciences, \\
National Taiwan University, Taipei 106, Taiwan}
\author{Jusak Tandean}
\email{jtandean@yahoo.com}
\affiliation{Department of Physics and Center for Theoretical Sciences, \\
National Taiwan University, Taipei 106, Taiwan}

\date{\today $\vphantom{\bigg|_{\bigg|}^|}$}

\begin{abstract}

In light of recent experimental information from the CLEO, BaBar, KTeV, and Belle collaborations,
we investigate some consequences of the possibility that a light spin-one particle is responsible
for the three \,$\Sigma^+\to p\mu^+\mu^-$\, events observed by the HyperCP experiment.
In particular, allowing the new particle to have both vector and axial-vector couplings
to ordinary fermions, we systematically study its contributions to various processes involving
$b$-flavored mesons, including $B$-$\bar B$ mixing as well as leptonic, inclusive, and
exclusive $B$ decays. Using the latest experimental data, we extract bounds on its couplings
and subsequently estimate upper limits for the branching ratios of a number of $B$ decays with
the new particle.  This can serve to guide experimental searches for the particle in order
to help confirm or refute its existence.
\end{abstract}


\maketitle

\section{Introduction}

The detection of a new particle having a sub-GeV mass would likely hint at the presence of
physics beyond the standard model.
This possibility has been raised recently by the observation of three events for the rare
decay mode  \,$\Sigma^+\to p\mu^+\mu^-$\, with dimuon invariant masses narrowly
clustered around 214.3\,MeV by the HyperCP collaboration a few years ago~\cite{Park:2005ek}.
Although these events can be accounted for within the standard model (SM) when
long-distance contributions are properly included~\cite{Bergstrom:1987wr}, the probability
that the three events have the same dimuon mass in the SM is less than 1~percent.
This makes it reasonable to speculate that a light neutral particle, $X$,
is responsible for the observed dimuon-mass distribution via
the decay chain  \,$\Sigma^+\to p X\to p\mu^+\mu^-$\,~\cite{Park:2005ek}.

The new-particle interpretation of the HyperCP result has been theoretically explored to
some extent in the literature~\cite{He:2005we,Deshpande:2005mb,Gorbunov:2005nu,
He:2006uu,He:2006fr,Zhu:2006zv,Tatischeff:2007dz,Chen:2006xja,Chen:2007uv}.
Various ideas that have been proposed include the possibility that $X$ is spinless or that
it has spin one.
In the spinless case, $X$ could be a sgoldstino in supersymmetric
models~\cite{Gorbunov:2005nu} or a $CP$-odd Higgs boson in the next-to-minimal
supersymmetric standard model (NMSSM)~\cite{He:2006fr,Zhu:2006zv}.
In the case of $X$ being a spin-1 particle, one possible candidate is the gauge ($U$) boson
of an extra U(1) gauge group in some extensions of the~SM~\cite{Chen:2007uv}.

The presence of $X$ in \,$\Sigma^+\to p\mu^+\mu^-$\, implies that it also contributes to other
\,$|\Delta S|=1$\, transitions, such as the kaon decays \,$K\to\pi\mu^+\mu^-$.\,
In general, the contributions of $X$ to \,$|\Delta S|=1$\, processes fall into two types.
The first one is induced by the flavor-changing (effective) couplings of $X$ to $d s$.
In addition to these two-quark contributions, there are so-called four-quark contributions
of $X$, which arise from the combined effects of the usual four-quark  \,$|\Delta S|=1$\,
operators in the SM and the flavor-conserving couplings of $X$ to quarks, as well as its
interactions with the SM gauge fields~\cite{He:2006uu}.
Although the two-quark contributions are generally expected to dominate over the four-quark
ones, in some models the parameter space may have regions where the two types of
contributions are comparable in size and hence could interfere
destructively~\cite{He:2006uu,He:2006fr}.
Accordingly, to explore the $X$ hypothesis in detail and compare its predictions with
experimental results in a definite way,
it is necessary to work under some model-dependent assumptions.

There are a number experiments that have recently been performed or are still ongoing
to test the $X$ hypothesis~\cite{Tung:2008gd,Love:2008hs,ktev,e391a,belle}.
Their results have begun to restrict some of the proposed ideas on $X$ in the literature.
In particular, as already mentioned, $X$ could be a light $CP$-odd Higgs boson in the NMSSM.
In the specific NMSSM scenario considered in Ref.~\cite{He:2006fr}, $X$ does not couple
to up-type quarks and has the same flavor-conserving coupling $l_d$ to all down-type
quarks, implying that the four-quark contributions of $X$ to \,$|\Delta S|=1$\, decays are
proportional to~$l_d$~\cite{He:2006fr}.
Recent searches for the radiative decays \,$\Upsilon({\rm1S,2S,3S})\to\gamma X\to\gamma\mu^+\mu^-$\,
by the CLEO and BaBar collaborations~\cite{Love:2008hs} have come back negative and
imposed sufficiently small upper-bounds on $l_d$
to make the four-quark contributions negligible compared to the two-quark ones.
With only the two-quark contributions being present, the scalar part of the $sdX$ coupling is
already constrained by \,$K\to\pi\mu\mu$\, data to be negligibly small, whereas its pseudoscalar
part can be probed by \,$K\to\pi\pi\mu\mu$\, measurements~\cite{He:2005we,Deshpande:2005mb}.
There are now preliminary results on the branching ratio $\cal B$ of
\,$K_L\to\pi^0\pi^0X\to\pi^0\pi^0\mu^+\mu^-$\, reported by the KTeV and E391a
collaborations~\cite{ktev,e391a}.
The KTeV preliminary measurement \,${\cal B}<9.44\times10^{-11}$\, at 90\%~C.L.~\cite{ktev}
is the much more stringent of the two and has an upper bound almost 20 times smaller than
the lower limit \,${\cal B}_{\rm lo}=1.7\times10^{-9}$\, predicted in Ref.~\cite{He:2005we}
under the assumption that the $sdX$ pseudoscalar coupling, $g_P^{}$, is purely real.
However, there is a possibility that $g_P^{}$ has an imaginary part, and in the case where
this coupling is mostly imaginary the predicted lower bound, ${\cal B}_{\rm lo}$, can be much
smaller.\footnote{We gratefully acknowledge D. Gorbunov for pointing this out to us.}
More precisely, one can find that \,${\cal B}_{\rm lo}<7\times10^{-11}$,\, which evades
the above bound from KTeV, if  \,$|{\rm Im}\,g_P^{}|>0.98\,|g_P^{}|$\, and, moreover,
\,${\cal B}_{\rm lo}=1.7\times10^{-9}\,|\epsilon_K^{}|^2\sim8\times10^{-15}$\,
if $g_P^{}$ is purely imaginary, \,$\epsilon_K^{}\sim{\cal O}(0.002)$\, being the usual
$CP$-violation parameter in kaon mixing.
If the KTeV preliminary result stands in their final report, then it will have imposed
a significant constraint on~$g_P^{}$, restricting it to be almost purely imaginary, for
the scenario in which $X$ has spin zero and its four-quark contributions to flavor-changing
transitions are negligible.
To place stronger restrictions on~$g_P^{}$, it is important to look for the decays of particles
other than neutral kaons, such as \,$K^\pm\to\pi^\pm\pi^0X$\, and
\,$\Omega^-\to\Xi^-X$\,~\cite{Kaplan:2007nn}.

Although the $X$ couplings in the \,$|\Delta S|=1$\, sector are in general independent of those
in the \,$|\Delta B|=1$\, sector, there is also new information from the latter sector that
seems compatible with the results of the $K_L$ measurements.
Very recently the Belle collaboration has given a~preliminary report on their search for
a~spinless $X$ in \,$B\to\rho\mu^+\mu^-$\, and \,$B\to K^*\mu^+\mu^-$\, with
$m_{\mu\mu}^{}$ values restricted within a small region around \,$m_{\mu\mu}^{}=214.3$\,MeV.\,
They did not observe any event and provided stringent upper-bounds on the branching ratios of
\,$B\to\rho X$\, and \,$B\to K^*X$\,~\cite{belle}.

Unlike the spinless case, the scenario in which $X$ has spin one is not yet as strongly
challenged by experimental data, for it predicts that the lower limit of the branching ratio of
\,$K_L\to\pi^0\pi^0X\to\pi^0\pi^0\mu^+\mu^-$\, arising from the two-quark $dsX$ axial-vector
coupling, taken to be real, is \,$2\times10^{-11}$\,~\cite{He:2005we}.
This prediction is well below the preliminary upper-bound of \,$9.44\times10^{-11}$\, from
KTeV~\cite{ktev} and could get lower in the presence of an imaginary part of the $dsX$ coupling.
It is therefore interesting to explore the spin-1 case further, which we will do here.

In this paper we focus on the contributions of $X$ with spin~1 to a number of rare
processes involving mesons containing the $b$ quark.
We will not deal with specific models, but will instead adopt a model-independent
approach, assuming that $X$ has flavor-changing two-quark couplings to down-type quarks only
and that its four-quark contributions to flavor-changing transitions are negligible compared
to the two-quark ones.
Accordingly, since the $bdX$ and $bsX$ couplings generally are not related to the $sdX$
couplings, we further assume that the $b(d,s)X$ couplings each have both parity-even and
parity-odd parts, but we leave the parity of $X$ unspecified.
Specifically, we allow $X$ to have both vector and axial-vector couplings to $b(d,s)$.
The more limited case of $X$ being an axial-vector boson with only parity-even couplings
to $b(d,s)$ has been considered in Ref.~\cite{Chen:2006xja}.
Following earlier work~\cite{He:2005we}, to be consistent with HyperCP observations
we also assume that $X$ does not interact strongly and decays inside
the detector with \,${\cal B}(X\to\mu^+\mu^-)=1$.\,
In exploring the effect of $X$ with spin~1 on $B$ transitions, we will incorporate
the latest experimental information and obtain constraints on the flavor-changing
couplings of $X$ in order to predict upper bounds on the rates of a number of rare decays.
At this point it is worth pointing out that, since we let $X$ have vector couplings to $b(d,s)$,
the transitions in which we are interested include $B$ decays into $X$ and a pseudoscalar
meson, such as pion or kaon, which were not considered in Ref.~\cite{Chen:2006xja}.
As our numbers will show, most of the branching ratios of the decays we consider can be
large enough to be detected in near-future $B$ experiments.
This can serve to guide experimental searches for $X$ in order to help confirm or
rule out the spin-1 case.

\section{Interactions and amplitudes}

Assuming that $X$ has spin one and does not carry electric or color charge, we can express
the Lagrangian describing its effective couplings to a $b$~quark and
a light quark \,$q=d$ or $s$\, as
\begin{eqnarray} \label{LbqX}
{\cal L}_{bqX}^{} \,\,=\,\,
-\bar q\gamma_\mu^{}\bigl(g_{Vq}^{}-g_{Aq}^{}\gamma_5^{}\bigr)b\,X^\mu \,\,+\,\, {\rm H.c.}
\,\,=\,\,
-\bar q\gamma_\mu^{}\bigl(g_{{\rm L}q}^{}P_{\rm L}^{}+g_{{\rm R}q}^{}P_{\rm R}^{}\bigr)b\,X^\mu
\,\,+\,\, {\rm H.c.} ~,
\end{eqnarray}
where $g_{Vq}^{}$ and $g_{Aq}^{}$ parametrize the vector and axial-vector couplings, respectively,
\,$g_{{\rm L}q,{\rm R}q}^{}=g_{Vq}^{}\pm g_{Aq}^{}$,\, and
\,$P_{\rm L,R}^{}=\frac{1}{2}(1\mp\gamma_5^{})$.\,
Generally, the constants $g_{Vq,Aq}^{}$ can be complex.
In the following, we derive the contributions of these two-quark interactions of $X$ to
the amplitudes for several processes involving $b$-flavored mesons.
As mentioned above, we follow here the scenario in which the four-quark flavor-changing
contributions of $X$ are negligible compared to the effects induced by~${\cal L}_{bqX}$.

The first transition we will consider is $B_q^0$-$\bar B_q^0$ mixing, which is characterized
by the physical mass-difference $\Delta M_q$ between the heavy and light mass-eigenstates
in the $B_q^0$-$\bar B_q^0$ system.
This observable is related to the matrix element $M_{12}^q$ for the mixing by
\,$\Delta M_q=2\,|M_{12}^{q}|$,\, where \,$M_{12}^q=M_{12}^{q,\rm SM}+M_{12}^{q,X}$\,
is obtained from the effective Hamiltonian ${\cal H}_{b\bar q\to\bar b q}$ for the SM plus
$X$-mediated contributions using
\,$2m_{B_q} M_{12}^q=\bigl\langle B_q^0\bigr|{\cal H}_{b\bar q\to\bar b q}
\bigl|\bar B_q^0\bigr\rangle$\,~\cite{Buchalla:1995vs}.

The SM part of $M_{12}^q$ is dominated by the top loop and given by~\cite{Buchalla:1995vs}
\begin{eqnarray}
M_{12}^{q,\rm SM} \,\,\simeq\,\, \frac{G_{\rm F}^2 m_W^2}{12\pi^2}\,f_{B_q}^2 m_{B_q}^{}\,
\eta_B^{} B_{B_q}^{}\, \bigl(V_{tb}^{}V_{tq}^*\bigr)^2\, S_0^{}\bigl(m_t^2/m_W^2\bigr) ~,
\end{eqnarray}
where $G_{\rm F}$ is the usual Fermi constant,  $f_{B_q}$ is the $B_q$ decay-constant,
$\eta_B^{}$ contains QCD corrections, $B_{B_q}$ is a~bag parameter, $V_{kl}$ are elements of
the Cabibbo-Kobayashi-Maskawa (CKM) matrix, and the loop function
\,$S_0^{}\bigl(m_t^2/m_W^2\bigr)\simeq2.4$.\,
To determine the $X$ contribution $M_{12}^{q,X}$, we derive the effective Hamiltonian
${\cal H}_{b\bar q\to \bar b q}^X$ from the amplitude for the tree-level transition
\,$b\bar q\to X^*\to\bar b q$\, calculated from~${\cal L}_{bqX}$.
Thus
\begin{eqnarray} \label{Hbq2bq}
{\cal H}_{b\bar q\to \bar b q}^X &=&
\frac{\bar q\gamma^\mu\bigl(g_{{\rm L}q}^{}P_{\rm L}^{}+g_{{\rm R}q}^{}P_{\rm R}^{}\bigr)b\,
      \bar q\gamma_\mu^{}\bigl(g_{{\rm L}q}^{}P_{\rm L}^{}+g_{{\rm R}q}^{}P_{\rm R}^{}\bigr)b}
{2\Bigl(m_X^2-m_{B_q}^2\Bigr)}
\nonumber \\ && +\,\,
\frac{\Bigl\{\bar q\Bigl[\bigl(g_{{\rm L}q}^{}m_q^{}-g_{{\rm R}q}^{}m_b^{}\bigr)P_{\rm L}^{} +
                  \bigl(g_{{\rm R}q}^{}m_q^{}-g_{{\rm L}q}^{}m_b^{}\bigr)P_{\rm R}^{}\Bigr]b\Bigr\}^2}
{2\Bigl(m_X^2-m_{B_q}^2\Bigr)m_X^2} ~,
\end{eqnarray}
where we have used in the denominators the approximation \,$p_X^2=m_{B_q}^2$\, appropriate
for the $B_q$ rest-frame and included an overall factor of 1/2 to account for
the products of two identical operators.
In evaluating the matrix element of this Hamiltonian at energy scales \,$\mu\sim m_b^{}$,\,
one needs to include the effect of QCD running from high energy scales which mixes
different operators. The resulting contribution of $X$ is
\begin{eqnarray}
M_{12}^{q,X} &=&
\frac{f_{B_q}^2\, m_{B_q}^{}}{3\bigl(m_X^2-m_{B_q}^2\bigr)} \Biggl[
\bigl(g_{Vq}^2+g_{Aq}^2\bigr) P_1^{\rm VLL} +
\frac{g_{Vq}^2\,\bigl(m_b^{}-m_q^{}\bigr)^2+g_{Aq}^2\,\bigl(m_b^{}+m_q^{}\bigr)^2}{m_X^2}\,
P_1^{\rm SLL}
\nonumber \\ && \hspace{16ex} +\,\, \bigl(g_{Vq}^2-g_{Aq}^2\bigr) P_1^{\rm LR}
+ \frac{g_{Vq}^2\,\bigl(m_b^{}-m_q^{}\bigr)^2-g_{Aq}^2\,\bigl(m_b^{}+m_q^{}\bigr)^2}{m_X^2}\,
P_2^{\rm LR} \Biggr] ~,
\end{eqnarray}
where \,$P_1^{\rm VLL}=\eta_1^{\rm VLL} B_1^{\rm VLL}$,\,
\,$P_1^{\rm SLL}=-\mbox{$\frac{5}{8}$}\, \eta_1^{\rm SLL} R_{B_q} B_1^{\rm SLL}$,\, and
\,$P_j^{\rm LR}=-\mbox{$\frac{1}{2}$}\, \eta_{1j}^{\rm LR} R_{B_q} B_1^{\rm LR}
                      + \mbox{$\frac{3}{4}$}\, \eta_{2j}^{\rm LR} R_{B_q} B_2^{\rm LR}$,\,
\,$j\,=\,1,2$\,~\cite{Buras:2001ra},
with the $\eta$'s denoting QCD-correction factors, the $B$'s being bag parameters defined by
the matrix elements
$\bigl\langle B_q^0\bigr|\bar q\gamma^\mu P_{\rm L}^{}b\,
\bar q\gamma_\mu^{}P_{\rm L}^{}b \bigl|\bar B_q^0\bigr\rangle =
\bigl\langle B_q^0\bigr|\bar q\gamma^\mu P_{\rm R}^{}b\,
\bar q\gamma_\mu^{}P_{\rm R}^{}b \bigl|\bar B_q^0\bigr\rangle =
\mbox{$\frac{2}{3}$} f_{B_q}^2 m_{B_q}^2 B_1^{\rm VLL}$,\,
\,$\bigl\langle B_q^0\bigr|\bar q P_{\rm L}^{}b\,\bar q P_{\rm L}^{}b\bigl|\bar B_q^0\bigr\rangle
= \bigl\langle B_q^0\bigr|\bar qP_{\rm R}^{}b\,\bar qP_{\rm R}^{}b\bigl|\bar B_q^0\bigr\rangle
= -\mbox{$\frac{5}{12}$} f_{B_q}^2 m_{B_q}^2 R_{B_q} B_1^{\rm SLL}$,\,
\,$\bigl\langle B_q^0\bigr|\bar q\gamma^\mu P_{\rm L}^{}b\,
\bar q\gamma_\mu^{}P_{\rm R}^{}b\bigl|\bar B_q^0\bigr\rangle =
-\mbox{$\frac{1}{3}$} f_{B_q}^2 m_{B_q}^2 R_{B_q} B_1^{\rm LR}$,\, and
\,$\bigl\langle B_q^0\bigr|\bar q P_{\rm L}^{}b\,\bar q P_{\rm R}^{}b\bigl|\bar B_q^0\bigr\rangle
=\mbox{$\frac{1}{2}$} f_{B_q}^2 m_{B_q}^2 R_{B_q} B_2^{\rm LR}$,\,
and \,$R_{B_q}=m_{B_q}^2/\bigl(m_b^{}+m_q^{}\bigr){}^2$.\,
Bounds on $g_{Vq}^{}$ and $g_{Aq}^{}$ can then be extracted from comparing the measured
and SM values of~$\Delta M_q$.

The second transition of interest is \,$B_q^0\to\mu^+\mu^-$,\, which receives a contribution
from \,$B_q^0\to X^*\to\mu^+\mu^-$.\,  To derive the amplitude for the latter, we need not only
${\cal L}_{bqX}$, but also the Lagrangian describing \,$X\to\mu^+\mu^-$.\,
Allowing the $X$ interaction with $\mu$ to have both parity-even and -odd parts,
we can write the latter Lagrangian as
\begin{eqnarray} \label{LlX}
{\cal L}_{\mu X}^{} \,\,=\,\,
\bar\mu\gamma_\alpha^{}\bigl(g_{V\mu}^{}+g_{A\mu}^{}\gamma_5^{}\bigr)\mu\,X^\alpha ~,
\end{eqnarray}
where $g_{V\mu}^{}$ and $g_{A\mu}^{}$ are coupling constants, which are real due to
the hermiticity of~${\cal L}_{\mu X}$.
Using the matrix elements
\,$\bigl\langle0\bigr|\bar q\gamma^\mu b\bigl|\bar B_q^0\bigr\rangle =
\bigl\langle0\bigr|\bar q b\bigl|\bar B_q^0\bigr\rangle = 0$,\,
\,$\bigl\langle0\bigr|\bar q\gamma^\mu\gamma_5^{}b\bigl|\bar B_q^0(p)\bigr\rangle =
-i f_{B_q}p^\mu$,\,  and
\,$\bigl\langle0\bigr|\bar q\gamma_5^{}b\bigl|\bar B_q^0\bigr\rangle =
i f_{B_q}m_{B_q}^2/\bigl(m_b^{}+m_q^{}\bigr)$,\,
we then arrive at
\begin{eqnarray}
{\cal M}\bigl(\bar B_q^0\to X\to\mu^+\mu^-\bigr) \,\,=\,\,
-\frac{2i f_{B_q}^{}\, g_{Aq}^{}\, g_{A\mu}^{}\, m_\mu^{}}{m_X^2}\, \bar\mu\gamma_5^{}\mu ~.
\end{eqnarray}
The resulting decay rate is
\begin{eqnarray} \label{rate_B2ll}
\Gamma\bigl(\bar B_q^0\to X\to\mu^+\mu^-\bigr) \,\,=\,\,
\frac{f_{B_q}^2\,\bigl|g_{Aq}^{}\,g_{A\mu}^{}\bigr|^2 m_\mu^2}{2\pi\,m_X^4}\,
\sqrt{m_{B_q}^2-4m_\mu^2} ~.
\end{eqnarray}
This implies that we need, in addition, the value of $g_{A\mu}^{}$, which can be estimated from
the contribution of ${\cal L}_{\mu X}$ in Eq.~(\ref{LlX}) at one-loop level to the anomalous
magnetic moment of the muon,~$a_{\mu}$.
We will determine $g_{A\mu}^{}$ in the next section.
Before moving on to other transitions, we note that from ${\cal L}_{\mu X}$ follows the decay rate
\begin{eqnarray} \label{rate_X2ll}
\Gamma\bigl(X\to\mu^+\mu^-\bigr) \,\,=\,\,
\frac{g_{V\mu}^2\, m_X^{}}{12\pi}
\Biggl(1+\frac{2 m_\mu^2}{m_X^2}\Biggr) \sqrt{1-\frac{4 m_\mu^2}{m_X^2}} \,+\,
\frac{g_{A\mu}^2\, m_X^{}}{12\pi}\,\Biggl(1-\frac{4 m_\mu^2}{m_X^2}\Biggr)^{\!3/2} ~.
\end{eqnarray}

The next process that can provide constraints on $g_{Vq}^{}$ and $g_{Aq}^{}$ is
the inclusive decay \,$b\to q\mu^+\mu^-$,\, to which \,$b\to q X$\,  can contribute.
From ${\cal L}_{bqX}$ above, it is straightforward to arrive at the inclusive decay rate
\begin{eqnarray} \label{rate_b2qX}
\Gamma(b\to q X) &=& \frac{|\bm{p}_X^{}|}{8\pi\,m_b^2 m_X^2}
\Bigl\{\bigl|g_{Vq}^{}\bigr|^2\Bigl[\bigl(m_b^{}+m_q^{}\bigr)^2+2 m_X^2\Bigr]
\Bigl[\bigl(m_b^{}-m_q^{}\bigr)^2-m_X^2\Bigr] \nonumber \\ && \hspace*{11ex} +\,\,
\bigl|g_{Aq}^{}\bigr|^2\Bigl[\bigl(m_b^{}-m_q^{}\bigr)^2+2 m_X^2\Bigr]
\Bigl[\bigl(m_b^{}+m_q^{}\bigr)^2-m_X^2\Bigr]\Bigr\} ~,
\end{eqnarray}
where $\bm{p}_X^{}$ is the 3-momentum of $X$ in the rest frame of~$b$.
One may probe the \,$b\to q X$\, contribution to \,$b\to q\mu^+\mu^-$\, by examining
the measured partial rate of the latter for the smallest range available of the dimuon
mass, $m_{\mu\mu}^{}$, that contains \,$m_{\mu\mu}^{}=m_X^{}$.

We will also consider the exclusive decays  \,$B\to M X$,\, which contribute to
\,$B\to M\mu^+\mu^-$,\, where $M$ is a~pseudoscalar meson~$P$, scalar meson~$S$,
vector meson~$V$, or axial-vector meson~$A$.
To evaluate their decay amplitudes, we need the \,$\bar B\to M$\, matrix elements of the
\,$b\to q$ operators in ${\cal L}_{bqX}$.
The matrix elements relevant to \,$\bar B\to P X$\, and \,$\bar B\to S X$\, are
\begin{eqnarray}
\kappa_P^{}\,\bigl\langle P\bigl(p_P^{}\bigr)\bigr|\bar q\gamma^\mu b
\bigl|\bar B\bigl(p_B^{}\bigr) \bigr\rangle &\,=\,&
\frac{m_B^2-m_P^2}{k^2}\, k^\mu\, F_0^{B P} +
\Biggl[\bigl(p_B^{}+p_P^{}\bigr)^\mu-\frac{m_B^2-m_P^2}{k^2}\,k^\mu\Biggr] F_1^{B P} ~,
\\
i \kappa_S^{}\,\bigl\langle S\bigl(p_S^{}\bigr)\bigr|\bar q\gamma^\mu \gamma_5^{} b
\bigl|\bar B\bigl(p_B^{}\bigr) \bigr\rangle &\,=\,&
\frac{m_B^2-m_S^2}{k^2}\, k^\mu\,F_0^{BS} +
\Biggl[ \bigl(p_B^{}+p_S^{}\bigr)^\mu - \frac{m_B^2-m_S^2}{k^2}\,k^\mu \Biggr] F_1^{BS} ~,
\end{eqnarray}
and \,$\langle P|\bar q\gamma^\mu\gamma_5^{}b|\bar B\rangle =
\langle S|\bar q\gamma^\mu b|\bar B\rangle=0$,\,
where \,$k=p_B^{}-p_{P,S}^{}$,\, the factor $\kappa_P^{}$ has a value of~1 for
\,$P=\pi^-,\bar K,D$\, or \,$-\sqrt2$\, for \,$P=\pi^0$,\,
the values of $\kappa_S^{}$  will be given in the next section,
and the form factors $F_{0,1}^{BP,BS}$ each depend on~$k^2$.
For \,$\bar B\to V X$\, and \,$\bar B\to A X$,\, we need
\begin{eqnarray}
\kappa_V^{}\,\bigl\langle V\bigl(p_V^{}\bigr)\bigr|\bar q\gamma_\mu^{}b
\bigl|\bar B\bigl(p_B^{}\bigr)\bigr\rangle \,&=&\,
\frac{2 V^{B V}}{m_B^{}+m_V^{}}\, \epsilon_{\mu\nu\sigma\tau}^{}\,
\varepsilon_V^{*\nu} p_B^\sigma\, p_V^\tau ~, \\
\kappa_V^{}\,\bigl\langle V\bigl(p_V^{}\bigr)\bigr|\bar q\gamma^\mu\gamma_5^{}b
\bigl|\bar B\bigl(p_B^{}\bigr)\bigr\rangle  \,&=&\,
2 i A_0^{B V} m_V^{}\, \frac{\varepsilon_V^{*}\!\cdot\!k}{k^2}\, k^\mu
+ i A_1^{B V}\, \bigl( m_B^{}+m_V^{} \bigr) \biggl(
\varepsilon_V^{*\mu}-\frac{\varepsilon_V^{*}\!\cdot\!k}{k^2}\, k^\mu \biggr)
\nonumber \\ && - \,\,
\frac{i A_2^{B V}\, \varepsilon_V^{*}\!\cdot\!k}{m_B^{}+m_V^{}}
\biggl( p_B^\mu + p_V^\mu - \frac{m_B^2-m_V^2}{k^2}\, k^\mu \biggr) ~,
\end{eqnarray}
\begin{eqnarray}
\kappa_A^{}\,\bigl\langle A\bigl(p_A^{}\bigr)\bigr|\bar q\gamma^\mu b
\bigl|\bar B\bigl(p_B^{}\bigr)\bigr\rangle  \,&=&\,
-2i V_0^{B A} m_A^{}\, \frac{\varepsilon_A^{*}\!\cdot\!k}{k^2}\, k^\mu
- iV_1^{B A}\, \bigl( m_B^{} -m_A^{} \bigr) \biggl(
\varepsilon_A^{*\mu}-\frac{\varepsilon_A^{*}\!\cdot\!k}{k^2}\, k^\mu \biggr)
\nonumber \\ && + \,\,
\frac{i V_2^{B A}\, \varepsilon_A^{*}\!\cdot\!k}{m_B^{} -m_A^{}}
\biggl( p_B^\mu + p_A^\mu - \frac{m_B^2 -m_A^2}{k^2}\, k^\mu \biggr) ~, \\
\kappa_A^{}\,\bigl\langle A\bigl(p_A^{}\bigr)\bigr|\bar q\gamma_\mu^{} \gamma_5 b
\bigl|\bar B\bigl(p_B^{}\bigr)\bigr\rangle \,&=&\,
\frac{-2 A^{BA}}{m_B^{} -m_A^{}}\, \epsilon_{\mu\nu\sigma\tau}^{}\,
\varepsilon_A^{*\nu} p_B^\sigma\, p_A^\tau ~,
\end{eqnarray}
where \,$k=p_B^{}-p_{V,A}^{}$,\, the factor $\kappa_V^{}$ has a magnitude of~1 for
\,$V=\rho^-,\bar K^*,\phi,D^*$\, or \,$\sqrt2$\, for \,$V=\rho^0,\omega$,\,
the values of $\kappa_A^{}$ will be given in the next section, and the form factors $V^{BV}$,
$A_{0,1,2}^{BV}$, $V_{0,1,2}^{BA}$, and $A^{BA}$ are all functions of~$k^2$.
Since $X$ has spin~1, its polarization $\varepsilon_X^{}$ and momentum $p_X^{}$ satisfy
the relation \,$\varepsilon_X^*\cdot p_X^{}=0$.\,
The amplitudes for \,$\bar B\to P X$\, and $\bar B \to S X$ are then
\begin{eqnarray} \label{M_B2PX}
{\cal M}(\bar B\to P X) &\,=\,& \frac{2\, g_{Vq}^{}}{\kappa_P^{}}\, F_1^{B P}\,
\varepsilon_X^*\!\cdot\!p_P^{} ~, \\ \label{M_B2SX}
{\cal M}(\bar B\to S X) &\,=\,&
\frac{2i\, g_{Aq}^{}}{\kappa_S^{}}\, F_1^{B S}\, \varepsilon_X^*\!\cdot\!p_S^{} ~,
\end{eqnarray}
leading to the decay rates
\begin{eqnarray} \label{rate_B2PX}
\Gamma (B \to P (S) X) \,\,=\,\, \frac{|\bm{p}_X^{}|^3}{2\pi\, \kappa_{P (S)}^2\, m_X^2}
\Bigl| g_{V (A) q}^{}\,F_1^{BP (S)} \Bigr|^2 ~,
\end{eqnarray}
where $\bm{p}_X^{}$ is the 3-momentum of $X$ in the rest frame of $B$.
For \,$\bar B\to V X$\, and \,$\bar B\to A X$,\, the amplitudes are
\begin{eqnarray} \label{M_B2VX}
{\cal M}(\bar B\to V X) \,&=&\,
-\frac{i g_{Aq}^{}}{\kappa_V^{}} \Biggl[
A_1^{BV}\, \bigl(m_B^{}+m_V^{}\bigr)\,\varepsilon_V^*\!\cdot\!\varepsilon_X^* \,-\,
\frac{2A_2^{BV}\, \varepsilon_V^*\!\cdot\!p_X^{}\,\varepsilon_X^*\!\cdot\!p_V^{}}
{m_B^{}+m_V^{}} \Biggr]
\nonumber \\ && +\,\,
\frac{2 g_{Vq}^{}\, V^{BV}}{\kappa_V^{}\,\bigl(m_B^{}+m_V^{}\bigr)}\,
\epsilon_{\mu\nu\sigma\tau}^{}\, \varepsilon_V^{*\mu}\varepsilon_X^{*\nu}p_V^\sigma\,p_X^\tau ~,
\end{eqnarray}
\begin{eqnarray} \label{M_B2AX}
{\cal M}(\bar B\to A X) \,&=&\,
-\frac{i g_{Vq}^{}}{\kappa_A^{}} \Biggl[
V_1^{BA}\, \bigl(m_B^{}-m_A^{}\bigr)\,\varepsilon_A^*\!\cdot\!\varepsilon_X^* \,-\,
\frac{2V_2^{BA}\, \varepsilon_A^*\!\cdot\!p_X^{}\,\varepsilon_X^*\!\cdot\!p_A^{}}
{m_B^{}-m_A^{}} \Biggr]
\nonumber \\ && +\,\,
\frac{2 g_{Aq}^{}\, A^{BA}}{\kappa_A^{}\,\bigl(m_B^{}-m_A^{}\bigr)}\, \epsilon_{\mu\nu\sigma\tau}^{}\,
\varepsilon_A^{*\mu}\varepsilon_X^{*\nu}p_A^\sigma\, p_X^\tau ~.
\end{eqnarray}
The corresponding decay rates can be conveniently written as~\cite{Kramer:1991xw}
\begin{eqnarray} \label{rate_B2VX}
\Gamma(B\to M'X) \,\,=\,\, \frac{|\bm{p}_X^{}|}{8\pi\, m_B^2}
\Bigl( \bigl|H_0^{M'}\bigr|^2 + \bigl|H_+^{M'}\bigr|^2 + \bigl|H_-^{M'}\bigr|^2 \Bigr) ~,
\end{eqnarray}
where  \,$M'=V$ or $A$,\, \,$H_0^{M'}=-a_{M'}^{}\,x_{M'}^{}-b_{M'}^{}\bigl(x_{M'}^2-1\bigr)$,\,
and \,$H_\pm^{M'}=a_{M'}^{}\pm c_{M'}\,\sqrt{x_{M'}^2-1}$,\,
with  \,$x_{M'}^{}=\bigl(m_B^2 - m_{M'}^2 - m_X^2\bigr)/\bigl(2 m_{M'}^{}m_X^{}\bigr)$,\,
\begin{eqnarray} \label{abcV}
a_V^{} &=& \frac{g_{Aq}^{}\, A_1^{BV}}{\kappa_V^{}}\bigl(m_B^{}+m_V^{}\bigr) ~, \hspace{5ex}
b_V^{} \,=\, \frac{-2 g_{Aq}^{}\, A_2^{BV} m_V^{} m_X^{}}{\kappa_V^{}\,\bigl(m_B^{}+m_V^{}\bigr)} ~,
\hspace{5ex}
c_V^{} \,=\, \frac{2 g_{Vq}^{}\, m_V^{} m_X^{} V^{BV}}{\kappa_V^{}\,\bigl(m_B^{}+m_V^{}\bigr)} ~,
~~~~~ \\ \label{abcA}
a_A^{} &=& \frac{g_{Vq}^{}\, V_1^{BA}}{\kappa_A^{}}\bigl(m_B^{}-m_A^{}\bigr) ~, \hspace{5ex}
b_A^{} \,=\, \frac{-2 g_{Vq}^{}\, V_2^{BA}m_A^{}m_X^{}}{\kappa_A^{}\,\bigl(m_B^{}-m_A^{}\bigr)} ~,
\hspace{5ex}
c_A^{} \,=\, \frac{2 g_{Aq}^{}\, m_A^{} m_X^{} A^{BA}}{\kappa_A^{}\,\bigl(m_B^{}-m_A^{}\bigr)} ~.
\end{eqnarray}

In the next section, we employ the expressions found above to extract constraints on
the couplings $g_{Vq}^{}$ and $g_{Aq}^{}$ from currently available experimental information.
We will subsequently use the results to predict upper bounds for the branching ratios of
a number of $B$ decays.

Before proceeding, we remark that we have not included in ${\cal L}_{bqX}$ in Eq.~(\ref{LbqX})
the possibility of dipole operators of the form
\,$\bar q\sigma^{\mu\nu}(1\pm\gamma_5^{})b\,\partial_\mu X_\nu$.\,
They would contribute to the processes dealt with above, except for \,$B_q^{}\to\mu^+\mu^-$.\,
However, we generally expect the effects of these operators to be suppressed compared to those
of ${\cal L}_{bqX}$ by a factor of order \,$p_X^{}/\Lambda\sim m_b^{}/\Lambda$,\,
with $\Lambda$ being a heavy mass representing the new-physics scale, if their contributions all
occur simultaneously.

\section{Numerical Analysis}

\subsection{Constraints from \boldmath$B_q$-$\bar B_q$ mixing}

As discussed in the preceding section, the $X$ contribution $M_{12}^{q,X}$ to $B_q$-$\bar B_q$
mixing is related to the observable \,$\Delta M_q=2\,|M_{12}^{q}|$,\, where
\,$M_{12}^q=M_{12}^{q,\rm SM}+M_{12}^{q,X}$.\,
The experimental value $\Delta M_q^{\rm exp}$ can then be expressed in terms of the SM
prediction $\Delta M_q^{\rm SM}$ as
\begin{eqnarray} \label{DM}
\Delta M_q^{\rm exp} \,\,=\,\, \Delta M_q^{\rm SM}\,\bigl|1+\delta_q^{}\bigr| \,\,,
\hspace{5ex} \delta_q^{} \,\,=\,\, \frac{M_{12}^{q,X}}{M_{12}^{q,\rm SM}} \,\,,
\end{eqnarray}
and so numerically they can lead to the allowed range of $\delta_q$,
from which we can extract the bounds on $g_{Vq,Aq}^{}$.
Thus, with \,$\Delta M_d^{\rm exp}=(0.507\pm0.005){\rm\,ps}^{-1}$\,~\cite{pdg} and
\,$\Delta M_d^{\rm SM}=\bigl(0.560^{+0.067}_{-0.076}\bigr){\rm\,ps}^{-1}$\,~\cite{ckmfit},
using the approximation \,$\bigl|1+\delta_d^{}\bigr|\simeq 1+{\rm Re}\,\delta_d$,\,
we can extract the $1 \sigma$ range
\begin{eqnarray} \label{delta_d}
-0.22 \,\,<\,\, {\rm Re}\,\delta_d^{} \,\,<\,\, +0.03 ~.
\end{eqnarray}
Similarly, \,$\Delta M_s^{\rm exp}=(17.77\pm0.12){\rm\,ps}^{-1}$\,~\cite{pdg} and
\,$\Delta M_s^{\rm SM}=(17.6^{+1.7}_{-1.8}){\rm\,ps}^{-1}$\,~\cite{ckmfit} translate into
\begin{eqnarray} \label{delta_s}
-0.09 \,\,<\,\, {\rm Re}\,\delta_s^{} \,\,<\,\, 0.11 ~.
\end{eqnarray}
To proceed, in addition to \,$m_X^{}=214.3$\,MeV,\, we use
\,$m_b^{}=4.4$\,GeV,\, \,$P_1^{\rm VLL}=0.84$,\, \,$P_1^{\rm SLL}=-1.47$,\,
\,$P_1^{\rm LR}=-1.62$,\, \,$P_2^{\rm LR}=2.46$\,~\cite{Buras:2001ra},
CKM parameters from Ref.~\cite{ckmfit},
\,$f_{B_d}=190$\,MeV,\, \,$f_{B_s}=228$\,MeV,\, \,$\eta_B^{}=0.551$,\, \,$B_{B_d}=1.17$,\,
and \,$B_{B_s}=1.23$\,~\cite{ckmfit,Buchalla:1995vs},
as well as meson masses from Ref.~\cite{pdg}.
Also, we will neglect $m_d^{}$ and $m_s^{}$ compared to $m_b^{}$.
It follows that for the ratio in Eq.~(\ref{DM})
\begin{eqnarray} \label{red}
{\rm Re}\,\delta_d^{} &=& \Bigl\{-4.4\,\Bigl[({\rm Re}\,g_{Vd}^{})^2-({\rm Im}\,g_{Vd}^{})^2\Bigr]
- 8.2\, ({\rm Re}\,g_{Vd}^{})({\rm Im}\,g_{Vd}^{}) \nonumber \\ && ~
+\, 17\,\Bigl[({\rm Re}\,g_{Ad}^{})^2-({\rm Im}\,g_{Ad}^{})^2\Bigr]
+ 33\,({\rm Re}\,g_{Ad}^{})({\rm Im}\,g_{Ad}^{}) \Bigr\}\times10^{12} ~,
\nonumber \\ \phantom{|^{\int^|}}
{\rm Re}\,\delta_s^{} &=& \Bigl\{-2.5\,\Bigl[({\rm Re}\,g_{Vs}^{})^2-({\rm Im}\,g_{Vs}^{})^2\Bigr]
+ 0.2\, ({\rm Re}\,g_{Vs}^{})({\rm Im}\,g_{Vs}^{}) \nonumber \\ && ~
+ 9.9\,\Bigl[({\rm Re}\,g_{As}^{})^2-({\rm Im}\,g_{As}^{})^2\Bigr]
- 0.7\, ({\rm Re}\,g_{As}^{})({\rm Im}\,g_{As}^{}) \Bigr\}\times10^{11} ~.
\end{eqnarray}

\begin{figure}[ht] \vspace*{3ex}
\includegraphics[height=2.5in,width=2.6in]{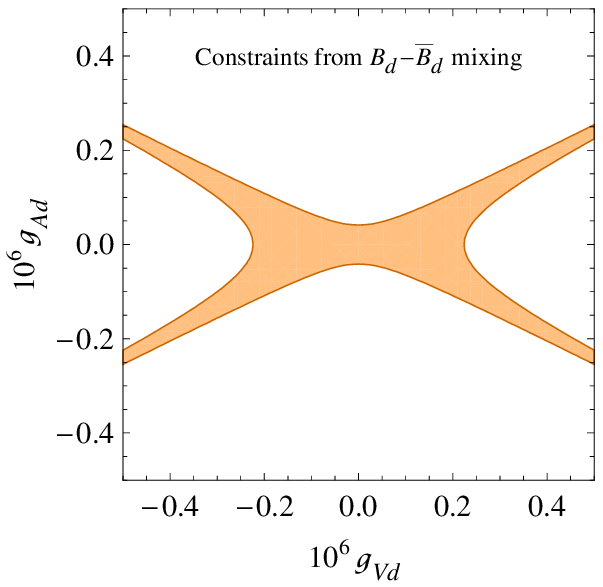} \hspace{5ex}
\includegraphics[width=2.5in,trim=0 0 0 0,clip]{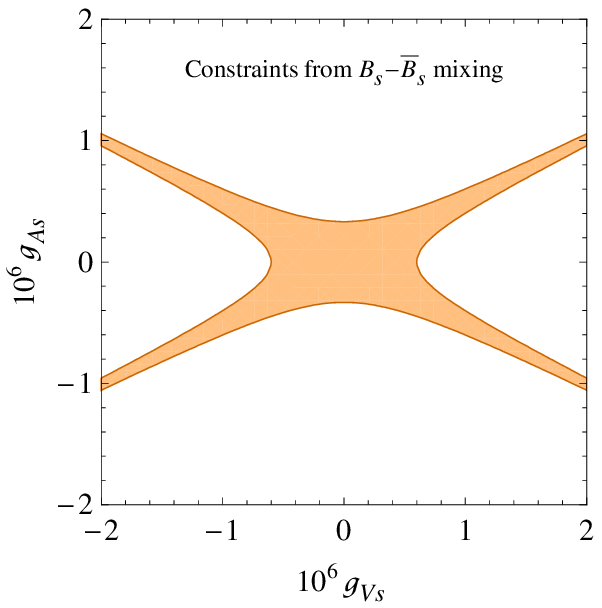} \vspace*{-1ex}
\caption{Parameter space of $g_{Vq}^{}$ and $g_{Aq}^{}$  subject to constraints from
$B_q$-$\bar B_q$ mixing, \,$q=d,s$,\, if $g_{Vq,Aq}^{}$ are taken to be real.\label{mix_bounds}}
\end{figure}

Hence constraints on the couplings come from combining these formulas
with Eqs.~(\ref{delta_d}) and~(\ref{delta_s}).
If only $g_{Vq}^{}$ or $g_{Aq}^{}$ contributes at a time, the resulting constraints are
\begin{eqnarray} \label{Bmix_bounds}
&& -0.7\times10^{-14} \,\,<\,\, ({\rm Re}\,g_{Vd}^{})^2-({\rm Im}\,g_{Vd}^{})^2
+ 1.9\, ({\rm Re}\,g_{Vd}^{})({\rm Im}\,g_{Vd}^{}) \,\,<\,\, 5.0\times10^{-14} ~,
\nonumber \\ \phantom{|^{\int}}
&& -1.3\times10^{-14} \,\,<\,\, ({\rm Re}\,g_{Ad}^{})^2-({\rm Im}\,g_{Ad}^{})^2
+ 1.9\, ({\rm Re}\,g_{Ad}^{})({\rm Im}\,g_{Ad}^{}) \,\,<\,\, 0.2\times10^{-14} ~,
\\ \phantom{|^{\int}}
&& -4.4\times10^{-13} \,\,<\,\, ({\rm Re}\,g_{Vs}^{})^2-({\rm Im}\,g_{Vs}^{})^2
- 0.1\, ({\rm Re}\,g_{Vs}^{})({\rm Im}\,g_{Vs}^{}) \,\,<\,\, 3.6\times10^{-13} ~,
\nonumber \\ \phantom{|^{\int}}
&& -0.9\times10^{-13} \,\,<\,\, ({\rm Re}\,g_{As}^{})^2-({\rm Im}\,g_{As}^{})^2
- 0.1\, ({\rm Re}\,g_{As}^{})({\rm Im}\,g_{As}^{}) \,\,<\,\, 1.1\times10^{-13} ~.
\end{eqnarray}
If one assumes instead that $g_{Vq,Aq}^{}$ are real, then from Eqs.~(\ref{delta_d})-(\ref{red})
one can determine the allowed ranges of the couplings shown in Fig.~\ref{mix_bounds}.

\subsection{Constraints from leptonic decays \,\boldmath$B_q\to\mu^+\mu^-$}

\begin{figure}[ht] \vspace*{2ex}
\includegraphics[width=2.5in]{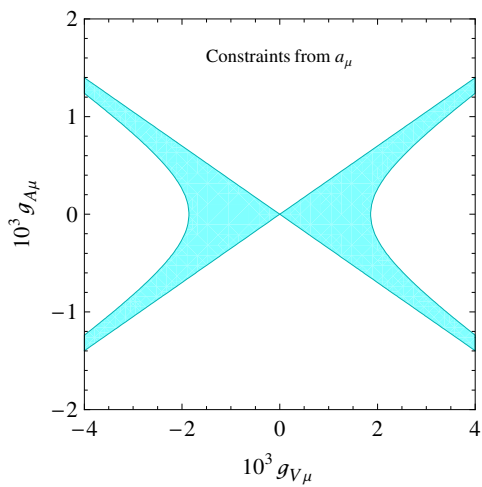} \vspace*{-1ex}
\caption{Parameter space of $g_{V\mu}^{}$ and $g_{A\mu}^{}$ subject to constraints from
the muon anomalous magnetic moment.\label{amu_bounds}} \vspace*{-3ex}
\end{figure}

As the \,$B_q\to\mu^+\mu^-$\, width in Eq.~(\ref{rate_B2ll}) indicates, to determine
$g_{Aq}^{}$ requires knowing the $X\mu\mu$ coupling constant~$g_{A\mu}^{}$.
Since ${\cal L}_{\mu X}$ in Eq.~(\ref{LlX}) generates the contribution of $X$ to
the muon anomalous magnetic moment~$a_\mu^{}$,  we may gain information on $g_{A\mu}^{}$
from~$a_\mu^{}$.
The $X$ contribution is calculated to be~\cite{He:2005we,Leveille:1977rc}
\begin{eqnarray} \label{amuX}
a_\mu^X \,\,=\,\,
\frac{m^2_\mu}{4\pi^2m^2_X}\bigl(g_{V\mu}^2\,f_V^{}(r)+g_{A\mu}^2\,f_A^{}(r)\bigr) \,\,=\,\,
1.1\times10^{-3}\,g_{V\mu}^2 \,-\, 9.0\times10^{-3}\,g_{A\mu}^2 ~,
\end{eqnarray}
where \,$r=m^2_\mu/m^2_X$,\,
\begin{eqnarray}
f_V^{}(r) \;=\; \int^1_0 dx\, \frac{x^2-x^3}{1-x +r x^2} ~, \hspace{5ex}
f_A^{}(r) \;=\; \int^1_0 dx\, \frac{-4 x+5 x^2-x^3-2r x^3}{1-x +r x^2} ~.
\end{eqnarray}
Presently there is a discrepancy of $3.2\sigma$ between the SM prediction for $a_\mu^{}$ and
its experimental value,
\,$\Delta a_\mu^{}=a_\mu^{\rm exp}-a_\mu^{\rm SM} =
(29\pm9)\times10^{-10}$\,~\cite{Jegerlehner:2009ry},
with \,$a_\mu^{\rm exp}=(11659208\pm6)\times 10^{-10}$\,~\cite{pdg}.
Consequently, since the $g_{V\mu}^{}$ and $g_{A\mu}^{}$ terms in $a_\mu^X$ are opposite in sign,
we require that \,$0<a_\mu^X<3.8\times10^{-9}$,\, which corresponds to the allowed parameter
space plotted in Fig.~\ref{amu_bounds}.
Avoiding tiny regions where the two terms in Eq.~(\ref{amuX}) have to conspire subtly to satisfy
the $a_\mu^X$ constraint, we then have
\begin{eqnarray}
g_{V\mu}^2 \,\,\lesssim\,\, 1\times10^{-5} ~, \hspace{5ex}
g_{A\mu}^2 \,\,\lesssim\,\, 1\times10^{-6} ~,
\end{eqnarray}
provided that \,$0<1.1\,g_{V\mu}^2-9.0\,g_{A\mu}^2<3.8\times10^{-6}$.\,
We note that combining these requirements for $g_{V\mu}^{}$ and $g_{A\mu}^{}$ with
Eq.~(\ref{rate_X2ll}) results in the width
\,$\Gamma\bigl(X\to\mu^+\mu^-\bigr)\lesssim1.8\times10^{-8}$\,GeV.\footnote{It is
worth mentioning here that in Ref.~\cite{He:2005we} the number for
\,$\Gamma\bigl(X_A\to\mu^+\mu^-\bigr)$\, in their Eq.~(18), corresponding to \,$g_{V\mu}^{}=0$\,
and  \,$g_{A\mu}^2=6.7\times10^{-8}$,\, is too large by a factor of~3.}\,
Assuming that \,$B_{d,s}\to X^*\to\mu^+\mu^-$\, saturates the latest measured bounds
\,${\cal B}\bigl(B_d\to\mu^+\mu^-\bigr)<6.0\times 10^{-9}$\, and
\,${\cal B}\bigl(B_s\to\mu^+\mu^-\bigr)<3.6\times 10^{-8}$\,~\cite{hfag}, respectively,
we use Eq.~(\ref{rate_B2ll}) with  \,$g_{A\mu}^2=1\times 10^{-6}$\, to extract
\begin{eqnarray} \label{B2ll_bounds}
|g_{Ad}^{}|^2 \,\,<\,\, 2.8\times10^{-14} ~, \hspace{5ex}
|g_{As}^{}|^2 \,\,<\,\, 1.2\times10^{-13} ~,
\end{eqnarray}
which are roughly comparable to the corresponding limits in Eq.~(\ref{Bmix_bounds})
from $B_q$-$\bar B_q$ mixing.

\subsection{Constraints from inclusive decay \,\boldmath$b\to q\mu^+\mu^-$}

Since there is still no experimental data on the inclusive \,$b\to d\mu^+\mu^-$,\,
we consider only the \,$q=s$\, case.
Thus, employing Eq.~(\ref{rate_b2qX}) and the $B_d^0$ lifetime~\cite{pdg}, we find
\begin{eqnarray} \label{rate_b2sX}
{\cal B}(b\to s X) \,\,\simeq\,\, \frac{\Gamma(b\to s X)}{\Gamma_{B_d^0}}
\,\,=\,\, 8.55\times10^{13}\,\bigl(|g_{Vs}^{}|^2+|g_{As}^{}|^2\bigr) \,\,.
\end{eqnarray}
To get constrains on $g_{Vs,As}^{}$, it is best to examine the measured partial rate for
the smallest $m_{\mu\mu}^{}$ bin available which contains \,$m_{\mu\mu}^{}=m_X^{}$.\,
The most recent data have been obtained by the BaBar and Belle
collaborations~\cite{Aubert:2004it,Iwasaki:2005sy}, the former giving the more restrictive
\begin{eqnarray}
{\cal B}(b\to s\ell^+\ell^-)_{m_{\ell\ell}^{}\in[0.2{\rm\,GeV},1.0{\rm\,GeV}]} \,\,=\,\,
\bigl(0.08\pm 0.36^{+0.07}_{-0.04}\bigr)\times 10^{-6}  ~,
\end{eqnarray}
which is the average over \,$\ell=e$ and~$\mu$.
This data allows us to demand that the $X$ contribution be below its 90\%-C.L.
upper-bound.
With \,${\cal B}(X\to\mu^+\mu^-)=1$,\, it follows that
\begin{eqnarray}
{\cal B}(b\to s X) \,\,<\,\, 6.8\times 10^{-7} ~,
\end{eqnarray}
which in combination with Eq.~(\ref{rate_b2sX}) implies
\begin{eqnarray} \label{incl_bound}
|g_{Vs}^{}|^2 \,+\, |g_{As}^{}|^2 \,\,<\,\, 8.0\times 10^{-21} ~.
\end{eqnarray}

\subsection{Constraints from exclusive decays \,\boldmath$B\to P\mu^+\mu^-$}

It can be seen from Eq.~(\ref{M_B2PX}) that only the vector coupling $g_{Vq}^{}$ is relevant to
the \,$B\to P X$ decay, not~$g_{Aq}^{}$.
As mentioned earlier, the possibility of $X$ having vector couplings was not considered in
Ref.~\cite{Chen:2006xja}, and therefore \,$B\to PX$\, decays were not studied therein.
Currently there is experimental information available on
\,$B\to\pi\mu^+\mu^-$\, and \,$B\to K\mu^+\mu^-$\, that can be used to place constraints
on~$g_{Vq}^{}$.  For the form factors $F_1^{BP}$, since they are functions of
\,$k^2=(p_B^{}-p_P^{})^2=m_X^2\ll m_B^2$,\,
it is a good approximation to take their values at \,$k^2=0$.\,
Thus, for \,$B\to(\pi,K)$\, we adopt those listed in Table~\ref{table1}.
Using Eq.~(\ref{rate_B2PX}), we then obtain
\begin{eqnarray} \label{rate_B2piX}
&& {\cal B} (B^+\to\pi^+ X) \,\,=\,\, 1.06\times10^{13}\, |g_{Vd}^{}|^2 ~,  \hspace{5ex}
{\cal B}\bigl(B_d \to\pi^0X\bigr) \,\,=\,\, 4.96\times10^{12}\, |g_{Vd}^{}|^2 ~, ~~~
\\ && \hspace*{10ex}
{\cal B}\bigl(B^+\to K^+X\bigr) \,\,\simeq\,\, {\cal B}\bigl(B_d \to K^0X\bigr) \,\,=\,\,
1.85\times10^{13}\, |g_{Vs}^{}|^2 ~. \phantom{|^{\int}} \label{rate_B2KX}
\end{eqnarray}
Experimentally, at present there are only upper limits for \,${\cal B}(B\to\pi\mu^+\mu^-)$,\,
namely~\cite{hfag,Wei:2008nv}
\begin{eqnarray}
{\cal B}(B^+\to\pi^+\mu^+\mu^-) \,\,<\,\, 6.9 \times 10^{-8} ~,  \hspace{5ex}
{\cal B}\bigl(B_d \to\pi^0\mu^+\mu^-\bigr) \,\,<\,\, 1.84 \times 10^{-7}
\end{eqnarray}
at 90\%~C.L.
Assuming that the contributions of \,$B\to\pi X\to\pi\mu^+\mu^-$\, saturate these bounds
and using Eq.~(\ref{rate_B2piX}) along with \,${\cal B}(X\to\mu^+\mu^-)=1$,\,
we find from the more stringent of them
\begin{eqnarray} \label{B2PX_bound}
|g_{Vd}^{}|^2 \,\,<\,\, 6.5 \times 10^{-21} ~.
\end{eqnarray}
For \,$B\to K\mu^+\mu^-$,\, there is data on the partial branching ratio that is pertinent
to \,$B\to K X$.\,  The latest measurement provides
\,${\cal B}(B\to K\mu^+\mu^-)_{m_{\mu\mu}\le2{\rm\,GeV}}=
\bigl(0.81^{+0.18}_{-0.16}\pm0.05\bigr)\times 10^{-7}$\,~\cite{Wei:2009zv}.
The corresponding SM prediction is consistent with this data~\cite{Antonelli:2009ws}
and has an uncertainty of about~30\%~\cite{Bobeth:2007dw}.  In view of this, we can
demand that \,${\cal B}(B\to K X\to K\mu^+\mu^-)$\,  be less than 40\% of the central
value of the measured result.\footnote{In estimating ${\cal B}(B\to M X\to M\mu^+\mu^-)$,
we neglect the interference in the \,$B\to M\mu^+\mu^-$\, rate between the SM and $X$
contributions because $X$ is very narrow, having a width of \,$\Gamma_X\lesssim10^{-8}$\,GeV,\,
as found earlier.}
Thus, with \,${\cal B}(X\to\mu^+\mu^-)=1$,\, we have
\begin{eqnarray}
{\cal B}(B\to K X) \,\,<\,\, 3.2\times 10^{-8}  ~.
\end{eqnarray}
Comparing this limit with Eq.~(\ref{rate_B2KX}) results in
\begin{eqnarray} \label{B2KX_bound}
|g_{Vs}^{}|^2 \,\,<\,\, 1.7\times 10^{-21} ~,
\end{eqnarray}
which is stronger than the $g_{Vs}^{}$ bound inferred from Eq.~(\ref{incl_bound}).
One can expect much better bounds on $g_{Vq}^{}$ from future measurements of
\,$B\to(\pi,K)\mu^+\mu^-$\, with $m_{\mu\mu}^{}$ values restricted within a small region
around \,$m_{\mu\mu}^{}=m_X^{}$.\,

\begin{table}[t]
\caption{Form factors relevant to \,$B\to P X$~\cite{Ball:2007hb}.} \smallskip
\begin{tabular}{|c|ccccccc|}
\hline
$$ & $\vphantom{\sum_|^|}\,$ $B_d\to\pi$ \, & \, $B_d\to\eta$ \, & \, $B_d\to\eta'$ \, &
\, $B_s\to K$ \, & \, $B_d\to K$ \, & \, $B_c\to D_d^+$ \, & \, $B_c\to D_s^+$ \, \\
\hline
\, $F_1^{BP}(0)\vphantom{\int_|^{|^|}}$ \, & 0.26 & 0.23 & 0.19 & 0.30 & 0.36 & 0.22 & 0.16 \\
\hline
\end{tabular} \label{table1} \medskip
\end{table}

\subsection{Constraints from exclusive decays \,\boldmath$B\to V\mu^+\mu^-$}

\begin{table}[b]
\caption{Form factors relevant to \,$B\to V X$~\cite{Ball:2004rg}.} \smallskip
\begin{tabular}{|c|ccccccc|}
\hline
\, $\vphantom{\sum_|^|}$\, & \, $B_d\to\rho$ \, & \, $B_d\to\omega$ \, & \, $B_s\to K^*$ \, &
\, $B_d\to K^*$ \, & \, $B_s\to\phi$ \, & \, $B_c\to D_d^{*+}$ \, & \, $B_c\to D_s^{*+}$ \, \\
\hline
\, $V^{BV}(0)\vphantom{\sum_|^|}$ \, & 0.32 & 0.29 & 0.31 & 0.41 & 0.43 & 0.63 & 0.54 \\
$A_1^{BV}(0)\vphantom{\sum_|^|}$     & 0.24 & 0.22 & 0.23 & 0.29 & 0.31 & 0.34 & 0.30 \\
$A_2^{BV}(0)\vphantom{\sum_|^|}$     & 0.22 & 0.20 & 0.18 & 0.26 & 0.23 & 0.41 & 0.36 \\
\hline
\end{tabular} \label{table2}
\end{table}

For \,$B\to V X$\, decays, the values of the relevant form factors at \,$k^2=0$\, are
listed in Table~\ref{table2}.
Employing those for \,$B=B_d$\, and \,$V=\rho,K^*$\, in Eq.~(\ref{rate_B2VX}), we find
\begin{eqnarray} \label{Br_rhoX1}
{\cal B}\bigl(B_d\to\rho^0 X\bigr) &=&
1.77 \times 10^{10}\,|g_{Vd}^{}|^2 + 6.18\times 10^{12}\,|g_{Ad}^{}|^2 ~, \nonumber \\
{\cal B}\bigl(B_d\to K^{*0}X\bigr) &=& \vphantom{|^{\int}}
5.45 \times 10^{10}\,|g_{Vs}^{}|^2 + 1.79\times 10^{13}\,|g_{As}^{}|^2 ~.
\end{eqnarray}
It is worth noting here that the dominance of the $g_{Aq}^{}$ terms in the preceding
formulas over the $g_{Vq}^{}$ terms also occurs in other \,$B\to V X$\, transitions and
corresponds to the fact that in the decay rate, Eq.~(\ref{rate_B2VX}), the $g_{Aq}^{}$ term
$|H_0^V|^2$ is significantly enhanced with respect to the $g_{Vq}^{}$ term
in~$|H_+^V|^2+|H_-^V|^2$.
Currently there is no published measurement of \,${\cal B}(B\to\rho\mu^+\mu^-)$,\,
but there are publicly available experimental data on \,${\cal B}(B\to K^*\mu^+\mu^-)$\,
for the $m_{\mu\mu}^{}$ bin containing \,$m_{\mu\mu}^{}=m_X^{}$,\, the most precise being
\,${\cal B}(B\to K^*\mu^+\mu^-)_{m_{\mu\mu}\le2\rm\,GeV}=
\bigl(1.46_{-0.35}^{+0.40}\pm0.11\bigr)\times10^{-7}$\,~\cite{Wei:2009zv}.
The corresponding SM prediction agrees with this data~\cite{Antonelli:2009ws}
and has an uncertainty of about~30\%~\cite{Beneke:2004dp}.
This suggests requiring \,${\cal B}(B\to K^*X\to K^*\mu^+\mu^-)$\, to be less than 40\% of
the central value of the measured result.
Thus, with \,${\cal B}(X\to\mu^+\mu^-)=1$,\, we have
\begin{eqnarray} \label{B_B2KX}
{\cal B}\bigl(B_d\to K^{*0} X\bigr) \,\,<\,\, 5.8\times10^{-8} ~.
\end{eqnarray}
In addition, very recently the Belle collaboration has provided a preliminary report on their
search for $X$ with spin~1 in \,$B\to\rho X\to\rho\mu^+\mu^-$\,  and
\,$B\to K^*X\to K^*\mu^+\mu^-$.\,
They did not observe any event and reported the preliminary bounds~\cite{belle}
\begin{eqnarray}
{\cal B}\bigl(B_d\to\rho^0X,\,\rho^0\to\pi^+\pi^-{\rm\,\,and\,\,}X\to\mu^+\mu^-\bigr) &<&
0.81\times 10^{-8} ~, \nonumber \\
{\cal B}\bigl(B_d\to K^{*0}X,\,K^{*0}\to K^+\pi^-{\rm\,\,and\,\,}X\to\mu^+\mu^-\bigr) &<&
1.53\times 10^{-8}  \vphantom{|^{\int}}
\end{eqnarray}
at 90\%~C.L.
Since \,${\cal B}(\rho^0\to\pi^+\pi^-)\simeq1$\, and \,${\cal B}(K^{*0}\to K^+\pi^-)\simeq2/3$,\,
these numbers translate into
\begin{eqnarray} \label{B_B2VX}
{\cal B}\bigl(B_d\to\rho^0X\bigr) \,\,<\,\, 0.81\times 10^{-8} ~, \hspace{5ex}
{\cal B}\bigl(B_d\to K^{*0}X\bigr) \,\,<\,\, 2.3\times 10^{-8} ~,
\end{eqnarray}
the second one being more restrictive than the constraint in Eq.~(\ref{B_B2KX}).
In the absence of more stringent limits, in the following we use these numbers inferred from
the preliminary Belle results.
Accordingly, applying the limits in Eq.~(\ref{B_B2VX}) to Eq.~(\ref{Br_rhoX1}) yields
\begin{eqnarray}  \label{B2rX_bound}
0.00286\,|g_{Vd}^{}|^2 \,+\, |g_{Ad}^{}|^2 &<& 1.3\times10^{-21} ~, \\ \label{B2KsX_bound}
0.00304\,|g_{Vs}^{}|^2 \,+\, |g_{As}^{}|^2 &<& 1.3\times10^{-21} ~. \vphantom{|^{\int}}
\end{eqnarray}
The $g_{As}^{}$ bound implied from the last equation can be seen to be stricter than that
from Eq.~(\ref{incl_bound}).

\begin{figure}[ht]
\includegraphics[height=2.5in,width=2.5in]{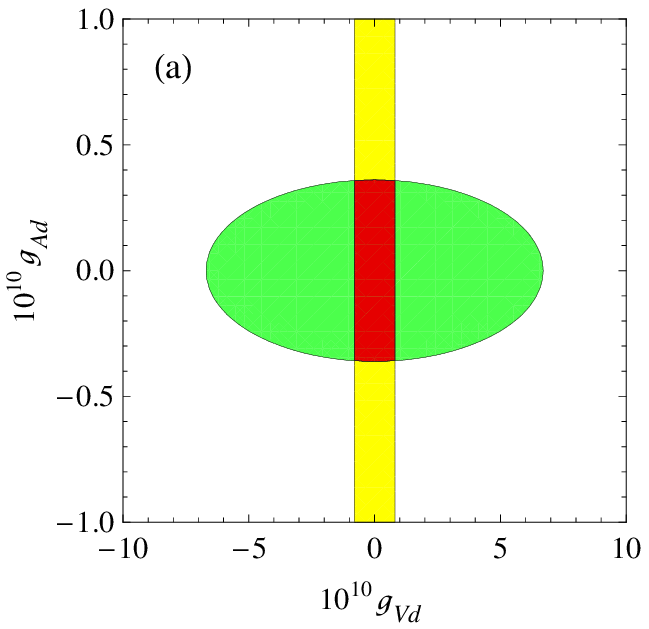} \hspace{5ex}
\includegraphics[height=2.5in,width=2.5in]{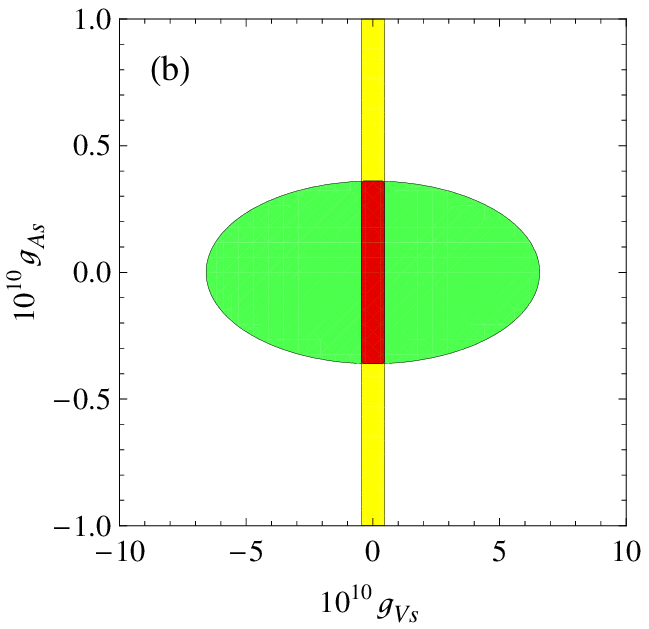} \vspace*{-1ex}
\caption{Parameter space of $g_{Vq}^{}$ and $g_{Aq}^{}$, taken to be real, subject to constraints on
(a)~$B\to\pi X$ (lightly shaded, yellow region), \,$B\to\rho X$ (medium shaded, green region),
and both of them (heavily shaded, red region) and
(b)~$B\to K X$ (lightly shaded, yellow region), \,$B\to K^*X$ (medium shaded, green region),
and both of them (heavily shaded, red region).\label{g_plots}}
\end{figure}

From Eqs.~(\ref{B2PX_bound}), (\ref{B2KX_bound}), (\ref{B2rX_bound}), and (\ref{B2KsX_bound}),
we can then extract the individual limits
\begin{eqnarray} \label{gd_bounds}
|g_{Vd}^{}|^2 \,\,<\,\, 6.5\times10^{-21} ~, & ~~~~~ &
|g_{Ad}^{}|^2 \,\,<\,\, 1.3\times10^{-21} ~, \\ \label{gs_bounds}
|g_{Vs}^{}|^2 \,\,<\,\, 1.7\times10^{-21} ~, & ~~~~~ &
|g_{As}^{}|^2 \,\,<\,\, 1.3\times10^{-21} ~. \vphantom{|^{\int}}
\end{eqnarray}
These bounds are clearly much stronger than those in Eqs.~(\ref{Bmix_bounds}) and~(\ref{B2ll_bounds})
derived from $B_q^0$-$\bar B_q^0$ mixing and \,$B_q^0\to\mu^+\mu^-$,\, respectively.
Also, combining Eqs.~(\ref{B2PX_bound}) and~(\ref{B2rX_bound}), we have plotted the allowed parameter
space of $g_{Vd}^{}$ and $g_{Ad}^{}$ in Fig.~\ref{g_plots}(a) under the assumption that they
are real.  Similarly, Fig.~\ref{g_plots}(b) shows the $g_{Vs}^{}$-$g_{As}^{}$ region allowed
by Eqs.~(\ref{B2KX_bound}) and~(\ref{B2KsX_bound}).

\subsection{Predictions for \,\boldmath$B\to MX$\, decays, \,$M=P,V,S,A$}

We can now use the results above to predict the upper limits for branching ratios of
a number of additional $B$-decays involving $X$.  Specifically, we explore two-body decays of
$B_{d,s}^0$ and $B_{u,c}^{}$ into $X$ and some of the lightest mesons~$M$.
We deal with \,$M=P$, $V$, $S$, and $A$\, in turn.

The $g_{Vd}^{}$ bound in Eq.~(\ref{gd_bounds}) leads directly to limits on
the branching ratios of \,$B_d^0\to\pi^0X$,\, \,$B_d^0\to\eta^{(\prime)}X$,\,
\,$B_s^0\to K^0X$,\, and \,$B_c^{}\to D_d^+X$.\,
Thus, from Eq.~(\ref{rate_B2piX}) follows
\begin{eqnarray}
{\cal B}\bigl(B_d^0\to\pi^0X\bigr) \,\,<\,\, 3.2\times10^{-8} ~.
\end{eqnarray}
Furthermore, employing Eq.~(\ref{rate_B2PX}) and Table~\ref{table1}, with
\,$\kappa_\eta^{}=\kappa_{\eta'}^{}=\sqrt2$,\, one gets
\begin{eqnarray}
{\cal B}\bigl(B_d^0\to\eta X\bigr) \,\,<\,\, 2.4\times10^{-8} ~, & ~~~~~ &
{\cal B}\bigl(B_d^0\to\eta'X\bigr) \,\,<\,\, 1.6\times10^{-8} ~, \nonumber \\
{\cal B}\bigl(B_s^0\to K^0X\bigr) \,\,<\,\, 8.2\times10^{-8} ~, & ~~~~~ &
{\cal B}\bigl(B_c^{}\to D_d^+X\bigr) \,\,<\,\, 1.7\times10^{-8} ~. \vphantom{|^{\int}}
\end{eqnarray}
Similarly, the $g_{Vs}^{}$ bound in Eq.~(\ref{gs_bounds}) implies
\begin{eqnarray}
&& {\cal B}\bigl(B_s^0\to\eta X\bigr) \,\,<\,\, 1.2\times10^{-8} ~, \hspace{5ex}
{\cal B}\bigl(B_s^0\to\eta'X\bigr) \,\,<\,\, 1.7\times10^{-8} ~, \nonumber \\ && \hspace*{15ex}
{\cal B}\bigl(B_c^{}\to D_s^+X\bigr) \,\,<\,\, 2.3\times10^{-9} ~, \vphantom{|^{\int}}
\end{eqnarray}
where the first two numbers have been calculated using \,$\kappa_\eta^{}=\kappa_{\eta'}^{}=1$,\,
\,$F_1^{B_s\eta}(0)=-F_1^{B_dK}(0)\,\sin\varphi$,\, and
\,$F_1^{B_s\eta'}(0)=F_1^{B_dK}(0)\,\cos\varphi$\,~\cite{Carlucci:2009gr}, with $F_1^{B_dK}(0)$
from Table~\ref{table1} and \,$\varphi=39.3^\circ$\,~\cite{Feldmann:1998vh}.

The $g_{Vq}^{}$ and $g_{Aq}^{}$ bounds in Eqs.~(\ref{gd_bounds}) and~(\ref{gs_bounds}),
together with Fig.~\ref{g_plots}, lead to upper limits for the branching ratios of
several other \,$B\to V X$\, decays.
Thus, combining Eq.~(\ref{rate_B2VX}) with the relevant form-factors in Table~\ref{table2}
yields for \,$q=d$\,
\begin{eqnarray}
{\cal B}(B^+\to\rho^+X) \,\,<\,\, 1.7\times 10^{-8}~, & ~~~~~ &
{\cal B}\bigl(B_d^0\to\omega X\bigr) \,\,<\,\, 7.0\times 10^{-9} ~, \nonumber \\
{\cal B}\bigl(B_s^0\to K^{*0}X\bigr) \,\,<\,\, 2.2\times 10^{-8} ~, & ~~~~~ &
{\cal B}\bigl(B_c^{}\to D_d^{*+}X\bigr) \,\,<\,\, 5.0\times 10^{-9} \vphantom{|^{\int}}
\end{eqnarray}
and for \,$q=s$\,
\begin{eqnarray}
{\cal B}\bigl(B_s^0\to\phi X\bigr) \,\,<\,\, 3.9\times 10^{-8} ~, \hspace{5ex}
{\cal B}\bigl(B_c^{}\to D_s^{*+}X\bigr) \,\,<\,\, 3.9\times 10^{-9} ~,
\end{eqnarray} \nopagebreak
where \,$|\phi\rangle\simeq|s\bar s\rangle$\, has been assumed.

In contrast to the \,$B\to P X$ case, $g_{Aq}^{}$ is the only coupling relevant to
\,$B\to SX$\, decays, as Eq.~(\ref{M_B2SX}) indicates.
From the $g_{Aq}^{}$ bounds found above, we can then estimate the branching ratios of some of
these decays.
Since the quark contents of many of the scalar mesons below 2\,GeV are not yet well established,
we consider only the decays with \,$S=a_0^{}(1450)$ and $K_0^*(1430)$,\, which are perhaps
the least controversial of the light scalar mesons~\cite{pdg}.
Adopting the form-factor values \,$F_1^{B_da_0(1450)}(0)=0.26$\, and
\,$F_1^{B_dK_0^*(1430)}(0)=0.26$\,~\cite{Cheng:2003sm}, we use Eq.~(\ref{rate_B2PX})
with \,$\kappa_S^{}=1$\, for \,$S=a_0^+(1450),K_0^*(1430)$\, and
\,$\kappa_S^{}=-\sqrt2$\, for \,$S=a_0^0(1450)$,\, as well as the $g_{Aq}^{}$ limits in
Eqs.~(\ref{gd_bounds}) and~(\ref{gs_bounds}), to obtain
\begin{eqnarray}
&& {\cal B}\bigl(B^+\to a_0^+(1450)X\bigr) \,\,<\,\, 1.1\times10^{-8} ~, \hspace{5ex}
{\cal B}\bigl(B_d^0\to a_0^0(1450)X\bigr) \,\,<\,\, 5.1\times10^{-9} ~, \nonumber \\ && \hspace*{5ex}
{\cal B}\bigl(B^+\to K_0^{*+}(1430)X\bigr) \,\,\simeq\,\,
{\cal B}\bigl(B_d^0\to K_0^{*0}(1430)X\bigr) \,\,<\,\, 1.0\times10^{-8} ~. \vphantom{|^{\int}}
\end{eqnarray}

Similarly to the \,$B\to V X$\, case, both $g_{Vq,Aq}^{}$ contribute to \,$B\to A X$,\,
as Eq.~(\ref{M_B2AX}) shows.
We will consider the decays with the lightest axial-vector mesons
\,$A=a_1^{}(1260)$, $b_1^{}(1235)$, $K_1(1270)$, and $K_1(1400)$.
The latter two are mixtures of the $K_{1A}$ and $K_{1B}$ states~\cite{pdg}, namely
\,$K_1(1270)=K_{1A}\,\sin\theta+K_{1B}\,\cos\theta$\, and
\,$K_1(1400)=K_{1A}\,\cos\theta-K_{1B}\,\sin\theta$,\, with \,$\theta=58^\circ$,\,
\,$m_{K_{1A}}=1.37$\,GeV,\, and \,$m_{K_{1B}}=1.31$\,GeV\,~\cite{Chen:2005cx}.
Incorporating the bounds in Eqs.~(\ref{gd_bounds}) and~(\ref{gs_bounds}) into
Eq.~(\ref{rate_B2VX}) with \,$\kappa_A^{}=1$\, for \,$S=a_1^+,b_1^+,K_1^{}$\, and
\,$\kappa_A^{}=-\sqrt2$\, for \,$S=a_1^0,b_1^0$,\, as well as the form factors listed
in Table~\ref{table3}, we arrive at
\begin{eqnarray} \label{B2AX}
{\cal B}\bigl(B^+\to a_1^+(1260)X\bigr) &\,\simeq\,&
2 {\cal B}\bigl(B_d^0\to a_1^0(1260)X\bigr) \,\,<\,\, 1.6\times10^{-8} ~, \nonumber \\
{\cal B}\bigl(B^+\to b_1^+(1235)X\bigr) &\,\simeq\,&
2 {\cal B}\bigl(B_d^0\to b_1^0(1235)X\bigr) \,\,<\,\, 1.2\times10^{-7} ~,  \vphantom{|^{\int}}
\nonumber \\
{\cal B}\bigl(B^+\to K_1^{*+}(1270)X\bigr) &\,\simeq\,&
{\cal B}\bigl(B_d^0\to K_1^{*0}(1270)X\bigr) \,\,<\,\, 2.6\times10^{-8} ~,  \vphantom{|^{\int}}
\nonumber \\
{\cal B}\bigl(B^+\to K_1^{*+}(1400)X\bigr) &\,\simeq\,&
{\cal B}\bigl(B_d^0\to K_1^{*0}(1400)X\bigr) \,\,<\,\, 1.3\times10^{-8} ~. \vphantom{|^{\int}}
\end{eqnarray}

\begin{table}[b]
\caption{Form factors relevant to \,$B\to A X$~\cite{Cheng:2003sm}.} \smallskip
\begin{tabular}{|c|cccc|}
\hline
\, $\vphantom{\sum_|^|}$\, & \, $B_d\to a_1^{}(1260)$ \, & \, $B_d\to b_1^{}(1235)$ \, &
\,\, $B_d\to K_{1A}$ \,\, & \,\, $B_d\to K_{1B}$ \,\,  \\
\hline
\, $A^{BA}(0)\vphantom{\sum_|^|}$ \, & 0.25 &  0.10  & 0.26 &  0.11 \\
$V_1^{BA}(0)\vphantom{\sum_|^|}$     & 0.37 &  0.18  & 0.39 &  0.19 \\
$V_2^{BA}(0)\vphantom{\sum_|^|}$     & 0.18 & $-$0.03~~ & 0.17 & $-$0.05~~ \\
\hline
\end{tabular} \label{table3}
\end{table}

Before ending this section, we would like to make a few more remarks regarding our results above.
The branching ratios of  \,$B^+\to\rho^+X$,\, \,$B_s^0\to\phi X$,\,
\,$B_d^0\to K_0^*(1430)X$,\, and \,$B\to K_1^{}X$\, were also estimated in
Ref.~\cite{Chen:2006xja} under the assumption that the vector couplings \,$g_{Vd,Vs}^{}=0$.\,
Compared to their numbers, our \,$B^+\to\rho^+X$\, result above is of similar order,
but our numbers for \,$B_s^0\to\phi X$\, and \,$B_d^{}\to K_0^*(1430)X$\, are smaller
by almost two orders of magnitude.  This is mostly due to the more recent data that we have
used to extract the $g_{Aq}^{}$ values.
On the other hand, our results for \,$B\to K_1^{}(1270)X,\,K_1^{}(1400)X$\, are larger
than the corresponding numbers in Ref.~\cite{Chen:2006xja} by up to two orders of magnitude.
The main cause of this enhancement is the nonzero contributions of~$g_{Vs}^{}$ to their decay
rates.  As one can see in Eq.~(\ref{rate_B2VX}) for the \,$B\to AX$\, rate, the $g_{Vq}^{}$
term in $|H_0^A|^2$ is significantly greater than the $g_{Aq}^{}$ term in
\,$|H_+^A|^2+|H_-^A|^2$.\,
For the same reason, without $g_{Vd}^{}$, the \,$B\to a_1^{}X,\,b_1^{}X$\, branching ratios
in Eq.~(\ref{B2AX}) would be orders of magnitude smaller.
Thus our inclusion of the vector couplings of $X$ has not only given rise to nonvanishing
\,$B\to PX$\, decays, but also helped make most of our predicted \,$B\to M X$\, branching
ratios as large as $10^{-8}$ to $10^{-7}$, which are within the reach of near-future
$B$ measurements.

\section{Conclusions}

Recent searches carried out by the CLEO, BaBar, E391a, KTeV, and Belle collaborations for
the HyperCP particle, $X$, have so far come back negative.
Furthermore, the new preliminary result from KTeV  has led to significant experimental
restrictions on the $sdX$ pseudoscalar coupling in the scenario where $X$
is a spinless particle and has negligible four-quark flavor-changing interactions.
In contrast, the possibility that $X$ is a spin-1 particle is not well challenged by
experiment yet.
In this paper, we have investigated some of the consequences of this latter possibility.
Specifically, taking a~model-independent approach, we have allowed $X$ to have both
vector and axial-vector couplings to ordinary fermions.
Assuming that its four-quark flavor-changing contributions are not important compared to
its two-quark $bqX$ interactions, we have systematically studied the contributions of $X$
to various processes involving $b$-flavored mesons, including $B_q$-$\bar B_q$ mixing,
\,$B_q\to\mu^+\mu^-$,\, inclusive \,$b\to q\mu^+\mu^-$,\, and exclusive \,$B\to M\mu^+\mu^-$\,
decays, with \,$q=d,s$\, and $M$ being a spinless or \mbox{spin-1} meson.
Using the latest experimental data, we have extracted bounds on the couplings of $X$ and
subsequently predicted the branching ratios of a number of \,$B\to M X$\, decays, where $M$
is a~pseudoscalar, vector, scalar, or axial-vector meson.
The presence of the vector couplings $g_{Vq}^{}$ of $X$ has caused the decays with
a pseudoscalar $M$ to occur and also greatly enhanced the branching ratios of the decays
with an axial-vector~$M$.
The \,$B\to M X$\, branching ratios that we have estimated can reach the $10^{-7}$ level,
as in the cases of \,$B_s^0\to K^0X$\, and \,$B^+\to b_1^+(1235)X$,\,
which is comparable to the preliminary upper limits for the branching ratios of
\,$B_d\to\rho^0X,\,K^{*0}X$\, recently measured by Belle.
Therefore, we expect that the $B$ decays that we have considered here can be probed by
upcoming $B$ experiments, which may help confirm or rule out the new-particle interpretation
of the HyperCP result.

\acknowledgments

This work was supported in part by NSC and NCTS.
We thank Hwanbae Park and HyoJung Hyun for valuable discussions on experimental results.
We also thank X.G.~He and G.~Valencia for helpful comments.

\end{document}